\pdfoutput=1
\documentclass[10pt,journal]{IEEEtran}
\usepackage{cite}
\usepackage[caption=false]{subfig}
\usepackage{graphicx}                     
\graphicspath{ {Images/} }
\usepackage{hyperref}
\usepackage{amsmath}
\usepackage{enumitem}
\usepackage[T1]{fontenc}
\usepackage[utf8]{inputenc}
\usepackage{tabularx,ragged2e,booktabs,caption}
\usepackage{courier}
\usepackage{float}
\floatstyle{plaintop}
\restylefloat{table}

\begin{document}

\title{Optimum Reconfiguration of Routing Interconnection Network in APSoC Fabrics}

\author{Mostafa~Darvishi,~\IEEEmembership{Member~IEEE}

\thanks{M. Darvishi obtained the Ph.D. in Digital Microelectronics from Electrical Engineering Department of Polytechnique Montreal, QC, Canada, H3C 3A7. (e-mail: mostafa.darvishi@polymtl.ca).}}

%\markboth{Submitted to IEEE Transactions on Circuits and Systems II, July~2020}%

\maketitle

\begin{abstract}
This paper presents an automated algorithm for optimum configuration of routing interconnection network in Xilinx Zynq-7000 All programmable system-on-chip (APSoC) fabrics. A method to configure circuits with optimum routing resources is presented along with their performance parameters with and without the proposed algorithm. The proposed algorithm enables full control over routing resources for using different interconnection types in order to create routing-based circuit-under-test. The algorithm proposes the routing techniques through the 2-D array of switch matrices inside the interconnection network and automatically identifies the involved programmable interconnection points associated with a node. An experimental setup is proposed to measure the performance parameters such as slack time and power with and without the applied algorithm on the APSoC routing resources. The proposed setup requires no external equipment such as manufactured equipments or external instruments for performance measurement.
\end{abstract}


\IEEEpeerreviewmaketitle

\section{Introduction}

\IEEEPARstart{F}{ield} Programmable Gate Arrays (FPGAs) have attracted a lot of interest in various domains due to their high circuit density and growing performance capability. These semiconductor devices are structured in an array of configurable logic blocks (CLB) connected via a programmable routing interconnection network \cite{IEEEhowto:one, IEEEhowto:two}.
Advances in semiconductor technology enabled integrating programmable logics with complex systems in a single silicon die. These new Xilinx All Programmable System-on-Chip (APSoC) devices have been used extensively in different applications in recent years \cite{IEEEhowto:one, IEEEhowto:two, IEEEhowto:three, IEEEhowto:four, IEEEhowto:five}. The PL part of the APSoC is the FPGA itself and is used for implementation of different digital circuits and systems. The PS part contains a microcontroller. Hence, the functionality of any system implemented into an APSoC can be partitioned between PL and PS while the PS can also take the control over the PL. Among the APSoC devices, the Xilinx Zynq-7000 fabricated in the Taiwan Semiconductor Manufacturing Company’s 28 nm technology node has been vastly used for different applications in recent years \cite{IEEEhowto:six, IEEEhowto:seven, IEEEhowto:eight}. Routing resources in Xilinx APSoC fabrics are controlled by SRAM cells that are called configuration bits \cite{IEEEhowto:nine}.

This paper presents an automated algorithm for optimum configuring of the routing resources in the programmable logic (PL) of a Zynq-7000 APSoC device. We propose a method to configure circuits with optimum routing resources as well as the performance validation results with and without the proposed algorithm. The proposed algorithm enables full control over routing resources for using different interconnection types to create optimum routing paths for the implemented circuits. The automated algorithm is implemented in the Xilinx Vivado scripting tool. Also, the algorithm proposes a technique for traversing switch matrices (SM) inside the interconnection network and automatically identifies the involved programmable interconnection points (PIP) associated with an input or output pin of an SM. It is noted that the default routing optimizer of the Xilinx Vivado tool never considers the optimum routing paths for a targeted design for implementation if the timing constraints are met. Even in the case of timing violation, Vivado optimizer tool just informs the violatied paths while no solution than increasing the paths delays is proposed by default. We also propose an experimental setup to measure the performance parameters such as slack time and power with and without the applied algorithm. No external equipment (e.g., such as manufactured equipments or external instruments) is required for such measurements.

This paper is structured as follows. Some background information is presented in Section II. An overview of the routing resources in Zynq-7000 APSoC is presented in Section III. The proposed algorithm for routing resources in Zynq-7000 APSoC is presented in Section IV. Experimental setup and results are discussed in Section V. Conclusion and future works are finally drawn in Section VI.

\section{Background}

This paper focuses on developing and automated algorithm for optimum configuration of interconnection network in the PL resource in a Zynq-7000 device. This state-of-the-art APSoC device offers specialized modules merged with the PS in a single die. It is noted that almost 98\% of all memory elements in the PL part are configuration bits, of which more than 90\% control the routing resources \cite{IEEEhowto:one, IEEEhowto:five}.

Routing interconnection network in the PL section is configured through a 2-D array of SMs. An input pin of each SM comprises a set of PIPs where a PIP, a CMOS transistor switch, can be programmably turned on/off to add/remove interconnects throughout the network.

Recent studies confirms the vast involvement of routing interconnection network in a variety of applications such as clock tuning \cite{IEEEhowto:ten}, Time-to-Digital Converters (TDC) \cite{IEEEhowto:eleven, IEEEhowto:twelve}, Physical Unclonable Functions (PUF) \cite{IEEEhowto:thirteen}, and True Random Number Generators (TRNG) \cite{IEEEhowto:fourteen} have shown that full control of the routing path between two given points of the circuit is an essential requirement. It is noted that the placement of circuit elements can be fully controlled by the designer, while routing resources are less controllable.

\begin{figure*}
	\begin{center}
	\includegraphics[scale=0.55]{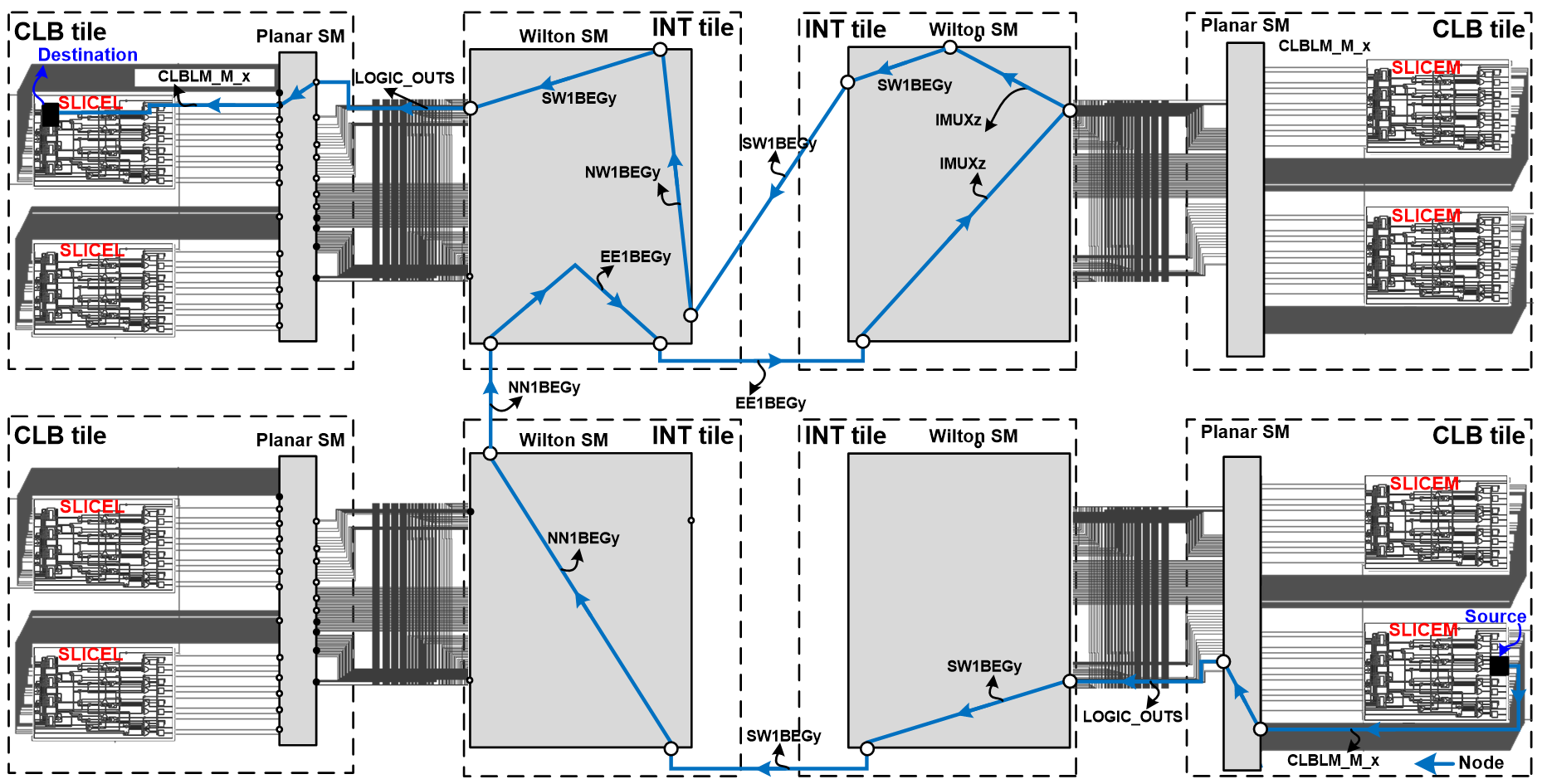}
	\caption{Topology of CLB and INT tiles in Xilinx APSoCs. In this example, Logical net  NetA connects a source flip-flop to a destination LUT made of seven nodes each made of SINGLE (1L) interconnects.}
	\label{fig:fig1}
	\end{center}
\end{figure*}

To make the proposed routing algorithm more comprehendible and also the ease of performance validation, the routing-based ring oscillators (RO) will be implemented as the circuits under test (CUT).

\section{OVERVIEW OF ROUTING RESOURCES IN ZYNQ-7000 APSOC}

Generic APSoC fabrics consist of some fundamental logic cores linked via an interconnection network. Three types of resources are commonly used in APSoC architectures: logic resources, routing interconnects, and switch matrices \cite{IEEEhowto:fifteen}. In this paper, we focus on the routing resources and switch matrices that are the roots of the interconnection network.

\subsection{Logic and Interconnect Tiles Resources}

Logic resources in APSoCs are linked via an interconnection network comprised of different interconnection types. Some interconnects are dedicated to specific logics or functions and the rests are global. Interconnects in the network span in both horizontal and vertical planes traversing the gate array from west to east and north to south, respectively. SMs are used to link various interconnects and transmit data inside the fabric.

Fig.~\ref{fig:fig1} shows the topology of a CLB and interconnect (INT) tiles in Xilinx 7-Series APSoCs \cite{IEEEhowto:three, IEEEhowto:five}. In this scheme, the planar SM has an injective mapping where each input node on the right side is connected to only one node on its left side. The INT tile comprises one Wilton SM (WSM) where each input node has a multiple mapping possibility to several output nodes and vice versa \cite{IEEEhowto:sixteen}. A net, such as NetA in Fig.~\ref{fig:fig1}, comprises a list of nodes and represents a logic net. A WSM input node sends data signal to several outgoing nodes (called downhill node) and one of the PIPs connected to an output node is configured to receive data signal from one of its multiple incoming nodes (called uphill nodes). A PIP specifies a configurable connection between an SM input and an SM output comprised of a programmable CMOS transistor as depicted in Fig.~\ref{fig:fig1} \cite{IEEEhowto:seventeen}.

\subsection{Interconnect Types Available in 7-Series FPGAs}
Xilinx APSoCs generally consist of fifteen types of interconnects for data signal transmission throughout the PL side of the fabric. It is noted that these interconnects are not valid for the PS side cause it is made of the multi-core microcontroller cores with dedicated interconnection network for itself and is out of the focus in this paper. Interconnects linking the Wilton SMs are categorized as follows \cite{IEEEhowto:three, IEEEhowto:five}:
\begin{itemize}
\item \textbf{SINGLE (1L)}: unidirectional interconnects that span 1 CLB;
\item \textbf{DOUBLE (2L)}: unidirectional interconnects that span 1 or 2 CLBs;
\item \textbf{HQUAD (4L)}: unidirectional interconnects that span 4 CLBs;
\item \textbf{VQUAD}: unidirectional interconnects that span 6 CLBs;
\item \textbf{BOUNCEACROSS}: unidirectional interconnects that span 1 CLB only vertically;
\item \textbf{VLONG}: bidirectional long interconnects that span 20 CLBs vertically;
\item \textbf{VLONG12}: bidirectional long interconnects that span 12 CLBs vertically;
\end{itemize}

		\begin{table}
		\centering
    \begin{tabular}{ | c | c | c |}
    \hline
    \textbf {Interconnect Type}  & \textbf {Number of Interconnects} \\& \textbf {Connected to each Wilton SM} \\ \hline
    DOUBLE  & 70   
    \\ \hline
    SINGLE  & 68 
		\\ \hline
    BOUNCEACROSS  & 17 \\ 
		\hline
		VLONG   & 3 \\ 
		\hline
		HLONG   & 3 \\ 
		\hline
		PINFEED   & 42 \\ 
		\hline
		OUTBOUND    & 24* \\ 
		\hline
		BOUNCEIN    & 9 \\ 
		\hline
		PINBOUNCE    & 16 \\ 
		\hline
		GLOBAL    & 12 \\ 
		\hline
		HQUAD    & 17 \\ 
		\hline
		BENTQUAD    & 34 \\ 
		\hline
		VQUAD    & 18 \\ 
		\hline
		VLONG12    & 2 \\ 
		\hline
		HVCCGNDOUT    & 2 \\ 
		\hline
    \end{tabular}
		\caption{TYPES AND NUMBER OF INTERCONNECTS LINKED TO EACH WILTON SM IN XILINX APSOCS}\label{tab1}
		\end{table}

\begin{itemize}
\item \textbf{HLONG}: bidirectional long interconnects that span 20 CLBs horizontally;
\item \textbf{GLOBAL}: homogeneous and unidirectional interconnects that span 20 CLBs vertically and are dedicated to route specific signals (clock, reset, enable, etc.);
\item \textbf{BENTQUAD}: unidirectional interconnects that bend and span 6 CLBs;
\item \textbf{PINFEED}: short interconnects that link Wilton SM to planar SM (coming into planar SM);
\item \textbf{OUTBOUND}: short interconnects that link planar SM to Wilton SM (outgoing from planar SM); some of them also span 1 CLB;
\item \textbf{BOUNCEIN}: short internal Wilton SM interconnects at some input nodes used to bounce signal;
\item \textbf{PINBOUNCE}: short internal Wilton SM interconnects at some output nodes used to bounce signal;
\item \textbf{HVCCGNDOUT}: GND and VCC interconnects to link Wilton SM nodes to logic ‘0’ or logic ‘1’, respectively.
\end{itemize}

Table\ref{tab1} provides the number of each interconnect’s type connected to a Wilton SM [3,5]. Four out of a total of 24 OUTBOUND interconnects span only one CLB, while the other rests link planar SMs to the Wilton SMs (* in Table I). Interconnects of a same type may have different topologies.

A logical net, for example NetA shown in Fig.~\ref{fig:fig1}, comprises a list of nodes 
\{Source\, CLBLM\_M\_A\, LOGIC\_OUTS2\, SW1BEG1\, SW1BEG1\, NN1BEG1\, NN1BEG1\, EE1BEG1\, EE1BEG1\, IMUX7\, IMUX7\, SW1BEG1\, SW1BEG1\, NW1BEG1\, SW1BEG1\, LOGIC\_OUTS2\, CLBLM\_M\_D6\, Destitation\}
connecting a flip-flop source to a LUT destination between two cross CLBs. NetA has seventeen nodes made of SINGLE (1L) interconnect that spans only 1SM.

\begin{figure}
	\includegraphics[scale=0.55]{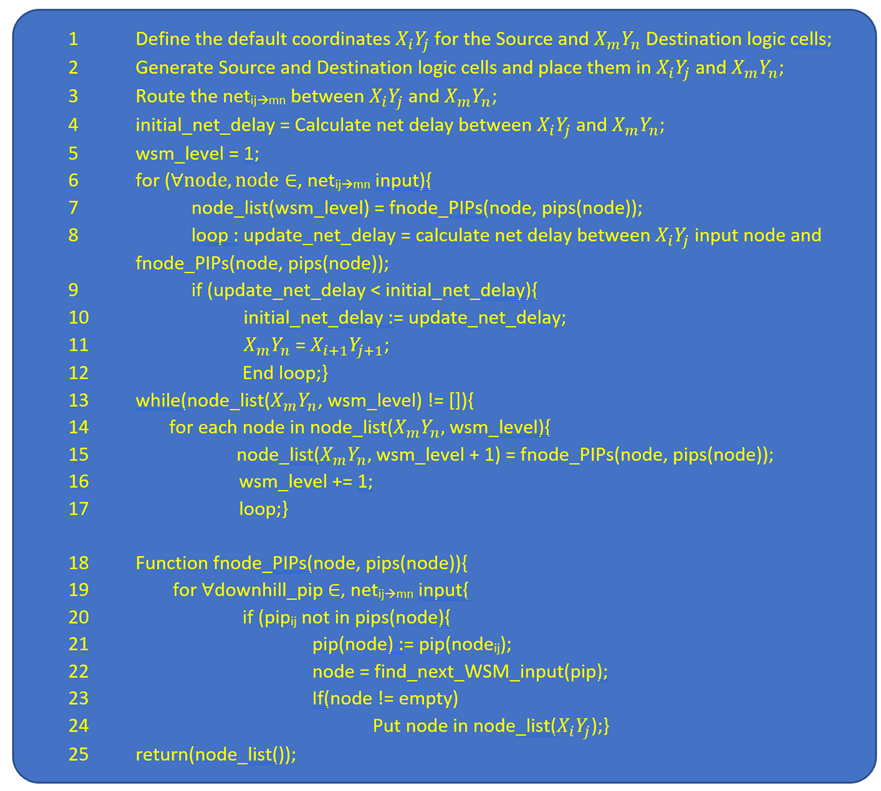}
	\caption{Proposed algorithm for WSMs.}
	\label{fig:fig2}
\end{figure}

\begin{figure}
	\includegraphics[scale=0.55]{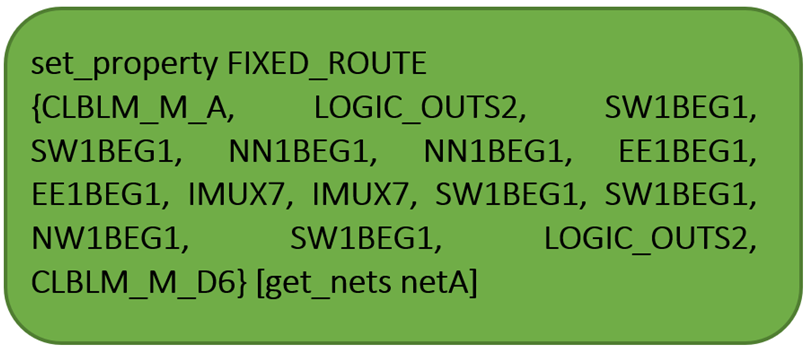}
	\caption{Proposed algorithm for WSMs.}
	\label{fig:fig3}
\end{figure}

\subsection{PIP Notation and Interconnect Coordinates}
In Zynq-7000 fabrics, usually a PIP is called by the name of interconnect it is connected to and the interconnect coordinates. The index of BEG or END is assigned to the PIP’s name depending on the interconnect’s tail being the beginning or the end of interconnect). For instance, a PIP in the tiles having coordinates X=5 and Y=15, connecting the beginning (BEG) of a SINGLE (1L) interconnect coming from southeast (NW) tile and the beginning (BEG) of a DOUBLE(2L) interconnect going to northwest (SE), is identified as:
\\*
\\*
\textbf{pip INT\_R\_X5Y15  NW1BEG0 ->  SE2BEG1}
\\*
\\*
The numbers before BEG introduces the interconnects’ length they are connecting (2 for DOUBLE interconnect and 1 for SINGLE interconnect in this example). The last number is an auto-assigned index by Vivado tool to distinguish the interconnects of a same category.

\section{PROPOSED ROUTING ALGORITHM FOR SWITCH MATRICES}

It is possible to determine all the PIPs associated with each node of \textbf{NetA} in Fig.~\ref{fig:fig1} that are connected to an input or output pin of a WSM. A trivial method is to select an input/output pin of a WSM and manually check the PIP Junction Properties

\begin{table}
    \begin{tabular}{ | c | c|}
    \hline
    \textbf {PIPs in wsm\_level=1:} & \textbf {PIPs in wsm\_level=2:} \\ \textbf {node connected to} & \textbf {node connected to}  \\ \textbf {LOGIC\_OUTS2}  & \textbf {NN1BEG3} \\ \hline
    WW4BEG0	,	NW2BEG0 & WW4BEG0	,	LV\_L0   
    \\ 
    WW2BEG0  , NR1BEG0 & WR1BEG1 , WW2BEG0
		\\ 
    WR1BEG1  , NN6BEG0 & NL1BEG\_N3 , WL1BEG2 \\ 
		
		WN1BEG\_N3 , NN1BEG3  & NW6BEG0 , SW6BEG3 \\ 
	
		SW6BE0  , NE6BEG0 & NW2BEG0 , SW2BEG3 \\ 

		SW2BEG0  , NE2BEG0 & NN6BEG0 , SS6BG3 \\ 

		SS6BEG0  , IMUX\_L8  & NN2BEG0 , SS2BEG3 \\ 

		SS2BEG0  , IMUX\_L40  & NE6BEG0 , SR1BEG1 \\ 
	
		SR1BEG1  , IMUX\_L32  & LV\_L18 , ER1BEG\_S0 \\ 

		SL1BEG0  , IMUX\_L24  & \\ 

		SE6BEG0  , IMUX\_L16   & \\ 

		SE2BEG0  , IMUX\_L0   & \\ 

		NL1BEG\_N3  , ER1BEG1  & \\ 

		BYP\_ALT0  , EL1BEG\_N3  & \\ 

		FAN\_ALT0  , EE4BEG0  & \\ 

		NW6BEG0 , EE2BEG0 & \\
		\hline
    \end{tabular}
		\caption{ EXAMPLE OF EXTRACTED PIPS CONNECTED TO LOGIC\_OUTS2 AND NN1BEG3 INTERCONNECTS IN WSM1 AND WSM4 RESULTED BY THE PROPOSED ALGORITHM}\label{tab2}
		\end{table}

in Vivado. This property identifies the PIPs of only one pin of a WSM at a time. It means that, it is not possible to identify the PIPs associated with several pins of a WSM, or the PIPs in different WSM levels simultaneously. This is very time consuming and inaccurate due to possible mistakes in net selection for designs including several nets. Automatic determination of all PIPs associated with each node facilitates the analysis of routing interconnection network in APSoC devices.

An algorithm is proposed for WSMs that automatically extracts all the PIPs in all WSM levels associated with each pin for any logic net as described in Fig.~\ref{fig:fig2}. In this pseudo-code, the source and destination logic cells are generated and placed with the FPGA floorplanner in STEP 1. The notion of logic cell here refers to a slice logic element, such as flip-flop or LUT. Then, a net is routed between the source and destination (STEP 2).

The proposed pseudo-code relies on the Xilinx Design Constraint (XDC) file, which provides full control of the placement and routing of a net in Vivado. Unlike using the automated place and route tasks performed by the tool, a group of nodes that the net should pass through can be specified. This can be achieved by employing the Tool Command Language (TCL) scripting available in Vivado. The FIXED\_ROUTE property allows the generation of a list of nodes to configure a net. This property should end with a specified name for the net that is going to be routed. For example, the NetA in Fig.~\ref{fig:fig1} is configured with the TCL script shown in Fig.~\ref{fig:fig3}. The indices associated with each node are pre-defined in Vivado to distinguish interconnects in the same category and cannot be changed. Table\ref{tab2} shows a partial result for the extracted PIPs connected to LOGIC\_OUTS2 and NN1BEG3 interconnects in WSM1 and WSM4 resulted by the proposed algorithm.

\begin{figure}
	\includegraphics[scale=0.55]{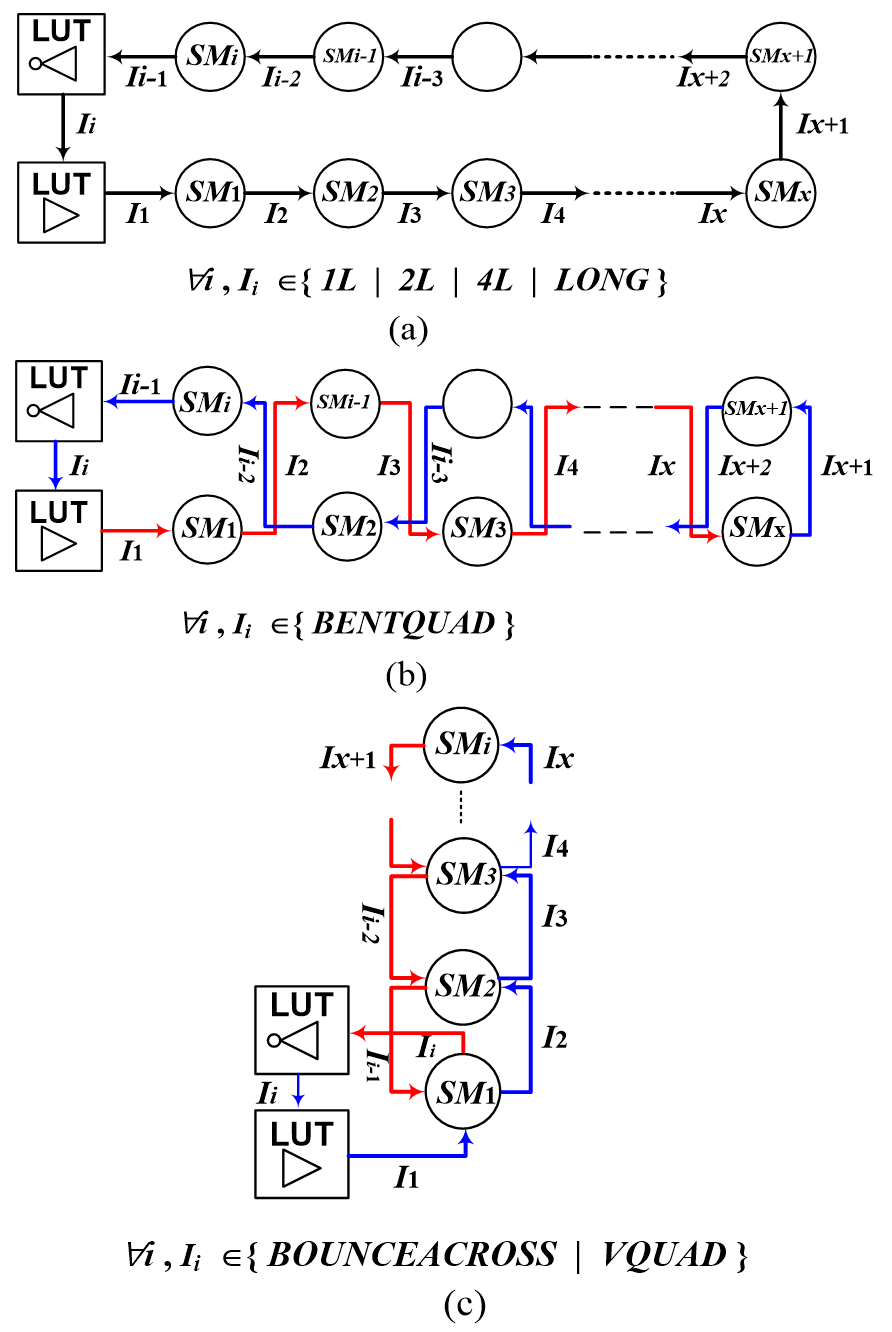}
	\caption{Routing diagram for three different sets of ROs preliminary implemented on the Xilinx APSoC “without” the proposed algorithm: (a) diagram of 1L, 2L, 4L and LONG interconnects, (b) diagram of BENTQUAD interconnects, and (c) diagram of BOUNCEACROSS and VQUAD interconnects \cite{IEEEhowto:three, IEEEhowto:five}.}
	\label{fig:fig4}
\end{figure}

Fig.~\ref{fig:fig4} shows a routing diagram representation of three different sets of ROs preliminary implemented on the Xilinx APSoC \cite{IEEEhowto:three, IEEEhowto:five} “without” using the proposed routing optimization algorithm. Fig.~\ref{fig:fig4}(a) to Fig.~\ref{fig:fig4}(c) show routing diagrams of horizontal ROs, the BENTQUAD RO, and the vertical ROs, respectively. In this figure, each Wilton SM is shown with a circle (SM) and each interconnect is shown with an arrow (I). The proposed RO architecture makes use of long routing paths and only two logic components.

\section {EXPERIMENTS AND RESULTS}

In the experiments, different ROs were implemented on the PL side of Zynq-7000 APSoC and their routing net delay as well as their frequencies were measured during run time.

\begin{figure}
	\includegraphics[scale=0.65]{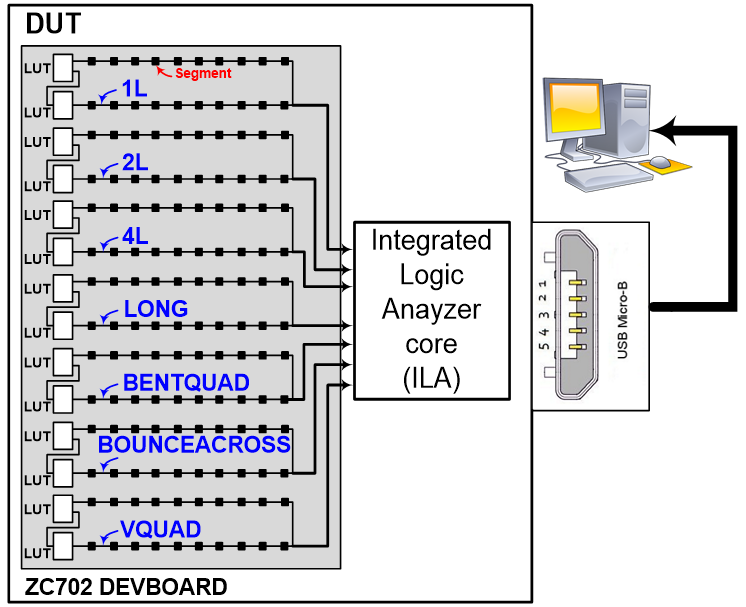}
	\caption{Experimental setup.}
	\label{fig:fig5}
\end{figure}

\begin{table}
\resizebox{\columnwidth}{!}{%
    \begin{tabular}{ | c | c | c | c |}
    \hline
    \textbf {RO Type} & \textbf {Frequncy} & \textbf {Net Delay} & \textbf {\# of} \\& \textbf {(kHz)} & \textbf {(ps)}  &\textbf {Interconnects}
		\\ \hline 
    1L	& 48912  & 398 & 51   \\
    1L  & 48909  & 402 & 52		\\ 
    2L  & 22541  & 696 & 56		\\
    2L  & 22541  & 696 & 56		\\
    4L  & 6399   & 183 & 60		\\
    4L  & 6398   & 182 & 60		\\
    LONG  & 16119  & 521 & 27		\\
    LONG  & 16121  & 516 & 26		\\
    BENTQUAD  & 23551  & 611 & 22		\\
    BENTQUAD  & 23548  & 615 & 23		\\
    BOUNCEACROSS  & 29852  & 489 & 25		\\
    BOUNCEACROSS  & 29851  & 490 & 25		\\
    VQUAD  & 29790  & 516 & 21		\\
    VQUAD  & 29789  & 519 & 22		\\
		\hline
    \end{tabular}
		\caption{PERFORMANCE RESULTS OF THE IMPLEMENTATION OF ROS USING THE PROPOSED ALGORITHM}\label{tab3}
		}
		\end{table}

\subsection {Setup and preliminary implementation}

The routing net delay and frequency of each individual RO is measured by using the using Xilinx Integrated Logic analyzer (ILA). Fig.~\ref{fig:fig5} shows the schematic of the implemented CUTs on the Zynq-7000 ZC702 APSoC using different types of interconnections as reported in Table I. It is noted that not all the interconnects reported in Table I are usable for RO implementation cause some of them do not span more than a single WSM (e.g. PINFEED interconnect). Two ROs per type were implemented that resulted in total of 14 ROs. Table \ref{tab3} shows the performance results of each RO “without” applying the proposed algorithm.

\subsection {Implementation with the Proposed Algorithm}

\begin{figure}
	\includegraphics[scale=0.55]{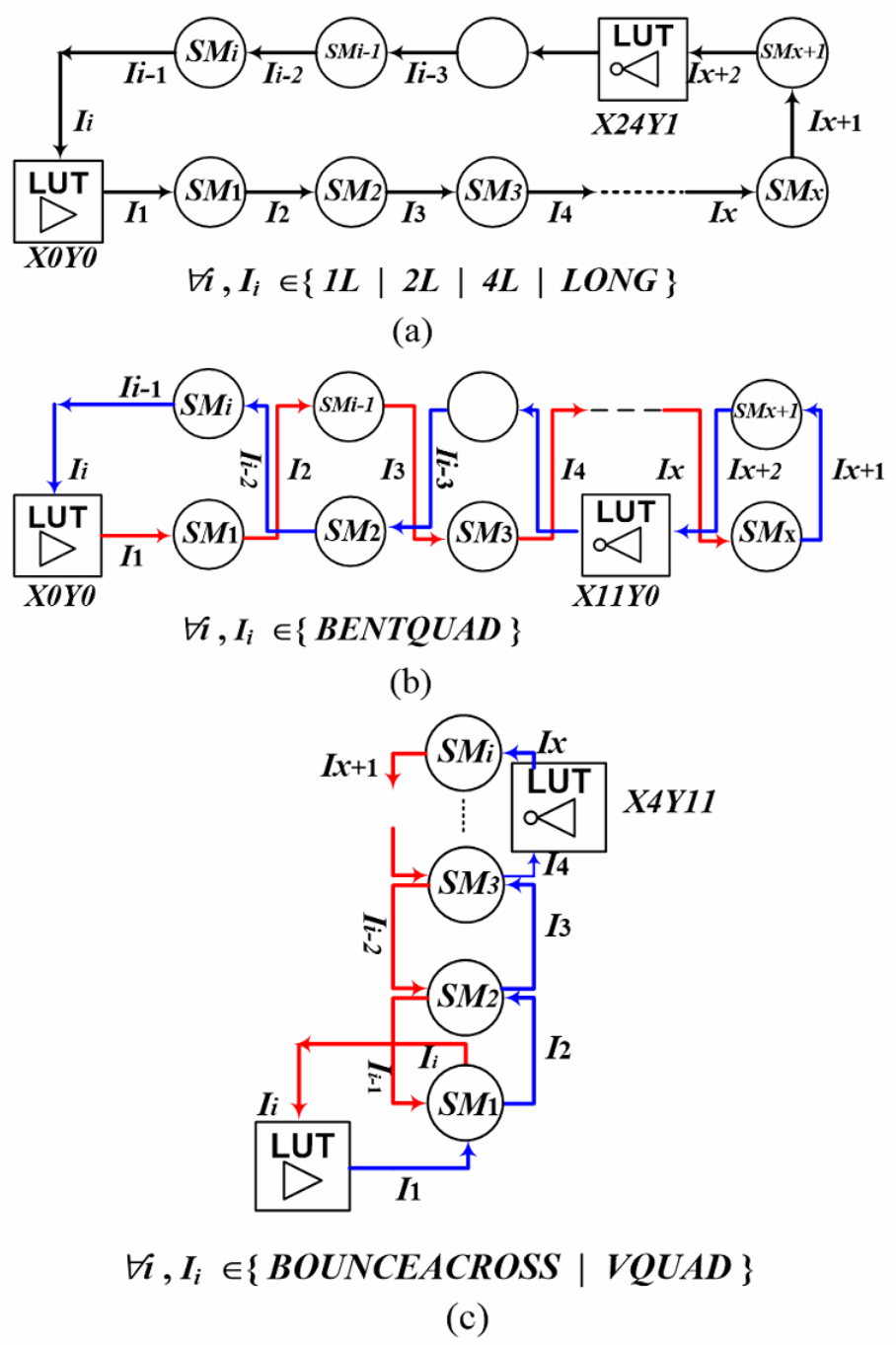}
	\caption{Routing diagram for three different sets of ROs implemented on the Xilinx APSoC “with” the proposed algorithm: (a) diagram of 1L, 2L, 4L and LONG interconnects, (b) diagram of BENTQUAD interconnects, and (c) diagram of BOUNCEACROSS and VQUAD interconnects.}
	\label{fig:fig6}
\end{figure}

In the next step, the Ros were implementation while the algorithm written in TCL script was also sourced during the design implementation process. Fig.~\ref{fig:fig6} shows updated routing diagram of the implemented ROs using the proposed algorithm. It is noted the optimized logic cell coordinates and routing topology has also been updated. Table IV shows the new measurements for the updated design using the proposed algorithm as shown in Fig.~\ref{fig:fig6}. It shows an improvement in the parameters specially the optimized number of utilized interconnects and the net delay.

\section {Conclusion}

This paper has presented a detailed analysis of an optimization algorithm applied to the routing resources in PL part of a Zynq-7000 APSoC that includes an SRAM-based FPGA (7Z020-CLG484) available on a ZC702. Fourteen ROs have been configured using two logic cells and routing resources of different interconnection types. The frequency and net delay of ROs has been measured using the ILA and delay measurement scripts in Vivado tool. The measurements have been performed for two scenarios: implementation of ROs “without” applying the proposed algorithm, and implementation “with” deploying the algorithm. In the former, the Vivado self-optimizer took care of the placement and routing while in the latter the proposed algorithm has overwritten the new placement and routing topology.

\end{document}